\documentclass[twocolumn,showpacs,amsmath,amssymb]{revtex4}

\usepackage{graphicx}
\usepackage{dcolumn}
\usepackage{bm}

\begin{document}

\preprint{APS/123-QED}

\title{Simultaneously optimizing the interdependent thermoelectric \\ parameters in Ce(Ni$_{1-x}$Cu$_x$)$_2$Al$_3$}

\author{Peijie Sun}
\altaffiliation[Present address: ]{Max Planck Institute for Chemical Physics of Solids, Dresden 01187, Germany.}
\author{Tusyoshi Ikeno}%
\author{Toshio Mizushima}
\author{Yosikazu Isikawa}
 \email[Corresponding author: ]{isikawa@sci.u-toyama.ac.jp}
\affiliation{%
Department of Physics, Toyama University, Toyama 930-8555, Japan}%

\date{\today}

\begin{abstract}
Substitution of Cu for Ni in the Kondo lattice system CeNi$_2$Al$_3$ results in a simultaneous optimization of the three interdependent thermoelectric parameters: thermoelectric power, 
electrical and thermal conductivities, where the electronic change in conduction band induced by the extra electron of Cu is shown to be crucial. The obtained thermoelectric figure of merit 
$zT$ amounts to 0.125 at around 100 K, comparable to the best values known for Kondo compounds. The realization of ideal thermoelectric optimization in Ce(Ni$_{1-x}$Cu$_x$)$_2$Al$_3$ 
indicates that proper electronic tuning of Kondo compounds is a promising approach to efficient thermoelectric materials for cryogenic application.  

\end{abstract}

\pacs{Valid PACS appear here}
\maketitle

Thermoelectric (TE) materials are used to construct all solid-state power generator or Peltier cooler. Current TE materials such as Bi$_2$Te$_3$-Sb$_2$Te$_3$ alloy have a dimensionless 
figure of merit $zT$$\sim$1 at near or above room temperature \cite{mahan2}. $zT$, defined as $T S^2$/$\rho \kappa$, determines the efficiency of TE conversion at a particular temperature 
$T$, with $S$ being the thermoelectric power, $\rho$ the electrical resistivity and $\kappa$ the thermal conductivity which consists of electronic part $\kappa _{\rm e}$ and lattice part 
$\kappa _{\rm L}$. In normal metals or semiconductors, $S$, $\rho$ and $\kappa$ are not independent. The classical route to high-efficiency TE materials is based on doping or alloying an 
appropriate semiconductor into a degenerate regime of carrier concentration $n$ $\sim$ 10$^{19}$ cm$^{-3}$, where a good compromise between $S$ and $\rho$ is reached with a high 
thermoelectric power factor $S^2/\rho$ \cite{mahan2}.
Modern electronic technology has created a strong demand for TE materials operating at cryogenic temperatures, to provide applications such as spot cooling microelectronic superconducting 
devices. The aforementioned concept, unfortunately, doesn't seem to hold true for development of TE materials below 200 K \cite{BiSb}. 

One promising candidate for cryogenic TE application is heavy-fermion (HF) or intermediate valence (IV) metals,
which are characterized by the so-called Kondo resonance of conduction electrons screening the magnetic moment of $f$ or $d$ electrons. Kondo resonance produces a sharp feature in the 
electronic density of states (DOS), leading to a 2$-$3 orders of magnitude enhancement of the effective charge-carrier mass $m^*$ and a large $S$ amounting to several tens of $\mu$V/K at 
around Kondo temperature $T_{\rm K}$ \cite{paschen}.  As metals, they have relatively large $\kappa _{\rm e}$ that is tied to metallic $\rho$ by Wiedeman-Franz law ($\rho\kappa _{e}$/$T$ = 
$L_0$, with $L_0$ = 2.44 $\times$ 10$^{-8}$ V$^2$/K$^2$).  A simple calculation neglecting $\kappa _{\rm L}$ indicates that $S$ $>$ 156 $\mu$V/K is necessary to achieve a practical $zT$ 
above unity. The largest $S$ so far known in metallic Kondo systems is $\sim$120 $\mu$V/K observed in CePd$_3$, with $zT$ $<$ 0.25 at around 200 K \cite{mahan2,phyT}. Further enhancement of 
$zT$ was not achieved for HF or IV systems, though a value as high as 14 has been predicted \cite{mahan1}.      

\begin{figure}[tp]
\includegraphics[width=0.85\linewidth]{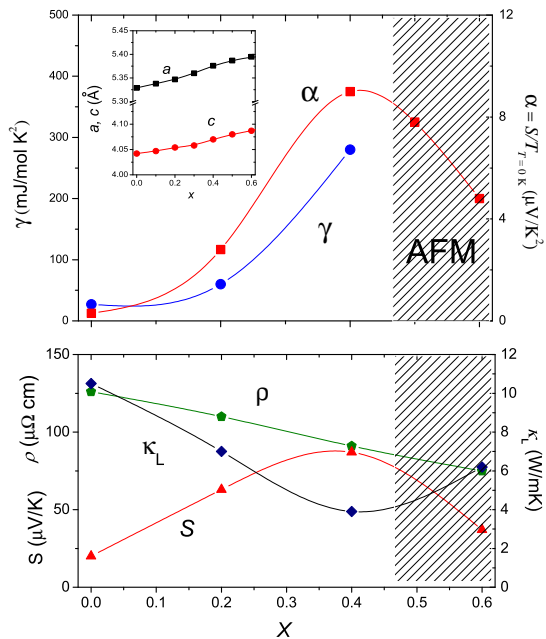}
\caption{Upper panel:  $x$ dependence of $\gamma$ and $\alpha$, the zero-temperature linear coefficients of electronic specific heat and thermoelectric power of 
Ce(Ni$_{1-x}$Cu$_x$)$_2$Al$_3$, respectively. The enhancement of both quantities with $x$ indicates the formation of heavy-fermion state with enhanced effective charge-carrier mass $m^*$.  
Lower panel: $x$ dependence  of the three thermoelectric parameters $S$, $\rho$ and $\kappa _L$ at $T$ = 100 K.  Their simultaneous optimization with increasing $x$ results in an enhanced 
thermoelectric figure of merit at $x$ = 0.4. For the compounds with $x$ $>$ 0.5, antiferromagnetic (AFM) order is confirmed. Inset shows  smooth increase of the lattice parameters $a$ and 
$c$ as a function of $x$.    
\label{gamma}}
\end{figure}

Like the current TE materials in use, high-$zT$ materials for cryogenic application might emerge by optimizing an appropriate parent system.  
The key point, therefore, is to identify this system and the mechanisms that might lead to optimization of the TE parameters.     
In this manuscript, we report a simultaneous optimization of $S$, $\rho$ and $\kappa _{\rm L}$ in a Kondo lattice system Ce(Ni$_{1-x}$Cu$_x$)$_2$Al$_3$ realized by substitution.     
The Ce ion in CeNi$_2$Al$_3$ experiences a nearly tetra-valent \cite{cava,isikawa} or IV state \cite{IV}, which, by replacing Ni with Cu, an atom with larger volume as well as an extra 
conduction electron, evolves into HF and eventually magnetically ordered state. The evolution accompanies an enhancement of the maximum thermoelectric power (Fig.~\ref{gamma}, lower panel) 
and its shift to lower temperature. Concomitant reductions of $\rho$ and $\kappa _{\rm L}$  due to the electronic change in conduction band and the chemical disorder, respectively, are also 
realized, resulting in a substantially enhanced $zT$ $\sim$ 0.125 at around 100 K. The largely enhanced electronic specific-heat coefficient $\gamma$ and $\alpha$ = $S/T$ ($T$ $\rightarrow$ 
0 K) (Fig.~\ref{gamma}, upper panel) point to the formation of heavy quasi-particles with substitution.   
At below 100 K, the observed $zT$ values largely exceed that of CePd$_3$, the so-called ``best" TE material in Kondo systems \cite{phyT}. The decisive impetus for the simultaneous 
optimization of $S$, $\rho$ and $\kappa _{\rm L}$ is to be discussed in terms of the electronic change of conduction band. 
A further development of more efficient TE materials for cryogenic application along the lines should be largely expected. 

Polycrystalline Ce(Ni$_{1-x}$Cu$_x$)$_2$Al$_3$ compounds with various $x$ were prepared by arc-melting stoichiometric amounts of constituent elements in a high purity argon atmosphere. The 
purities of the starting elements are 99.9\%, 99.997\%, 99.999\%, and 99.999\% for Ce, Ni, Cu and Al, respectively. 
No annealing treatment was carried out.
Power x-ray diffraction spectra confirmed the formation of the hexagonal PrNi$_2$Al$_3$-type structure of the samples with $x$ $<$ 0.5, with tiny impurity peak appearing at around 
2$\theta$$\sim$50$^\circ$ as reported previously \cite{Lu, cava}.  For $x$ $>$ $0.5$, a trace of second phase, which was determined to be CeCuAl$_3$ \cite{cecual}, appears gradually.  The 
second phase in the sample of $x$ = 0.6, i.e., the upper substitution limit in this work, was estimated to be less than 3\%. For this composition, a magnetic transition was observed at 2.3 K 
and it increases rapidly with $x$.  The continuous expansions of the lattice constants $a$ and $c$ (Fig.~\ref{gamma}, Inset) indicate successful substitution of Cu atoms. A change in the 
slopes  of $a$ and $c$ as a function of $x$ is discerned at around $x$ = 0.4, the critical concentration between the non-magnetic and magnetic regimes \cite{qct}. The thermoelectric 
properties were measured by either a home-build cryostat with a slowly-varying-gradient technique (chromel/Au+0.07\% Fe thermocouple as $\Delta T$ detector) or a commercial physical 
properties measurement system (PPMS, Quantum Design). The discrepancies between the two techniques were found in less than 5\% with a typical sample dimension of 1$\times$1$\times$5 mm$^3$. 

\begin{figure}[tp]
\includegraphics[width=0.85\linewidth]{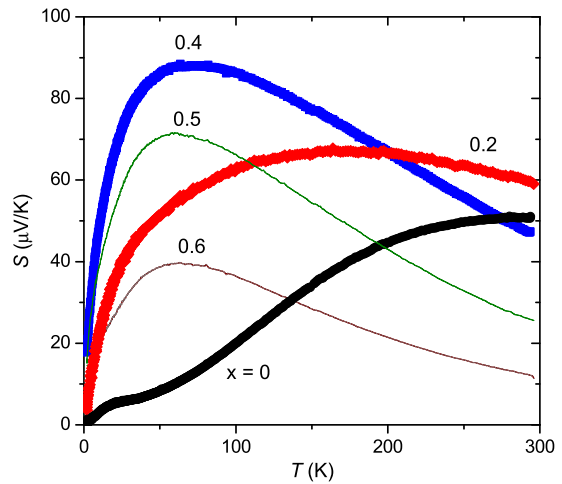}
\caption{The thermoelectric power $S$ as a function of temperature for various Ce(Ni$_{1-x}$Cu$_x$)$_2$Al$_3$ compounds. While the position of $S_{max}$ in temperature decreases with 
substitution up to $x$ = 0.4, its value increases largely. Above this critical concentration, the compounds order magnetically and their $S_{\rm max}$ decreases dramatically with $x$. 
\label{tep}}
\end{figure}

The most impressive feature in the $S(T)$ curves (Fig.~\ref{tep}) is a largely enhanced, positive maximum $S_{\rm max}$, as observed in many other Ce-contained HF or IV compounds 
\cite{mahan2}. For $x$ $\leq$ 0.4, while the position of $S_{\rm max}$ in temperature decreases, its value increases largely with $x$. The enhanced $S_{\rm max}$ value for the compound of 
$x$ = 0.4, 90 $\mu$V/K at 70 K, was confirmed by the two measurement techniques given above. The previous studies on annealed samples presented a $S_{\rm max}$ of only 60 $\mu$V/K for $x$ = 
0.4 \cite{Lu}. The difference with the present results most likely reflects the sensitivity of electron-electron correlations with respect to composition in the vicinity of the critical 
concentration: for example, $\gamma$ and $\alpha$ show dramatic change at around $x$ = 0.4 (Fig.~\ref{gamma}, upper panel). 
Looking carefully into the previous results \cite{Lu}, one also notice that the real $S_{\rm max}$ for $x$ = 0.4 could be higher: while $\gamma$ doubles by varying $x$ from 0.3 to 0.4, 
$\alpha$ does not follow $\gamma$ but remains flat. 
As estimated from the value of $1$/$\gamma$ and the position of $S_{\rm max}$, Kondo temperature $T_{\rm K}$ decreases gradually with $x$.
The enhancement of $S_{\rm max}$ with $x$ tracks the enhancement of the effective mass $m^*$ ($\sim$$\gamma$) as a consequence of the reduction of $T_{\rm K}$.   
The correspondence of the enhanced $\gamma$ and $\alpha$ (Fig.~\ref{gamma}, upper panel) follows the general behavior expected for correlated systems \cite{behnia, sakurai1}, unambiguously 
evidencing the realization of HF state. Further substitution exceeding $x$ = 0.4 reduces $S_{\rm max}$, due to the onset of magnetic ordering and the consequent diminish of $m^*$.

\begin{figure}[tp]
\includegraphics[width=0.85\linewidth]{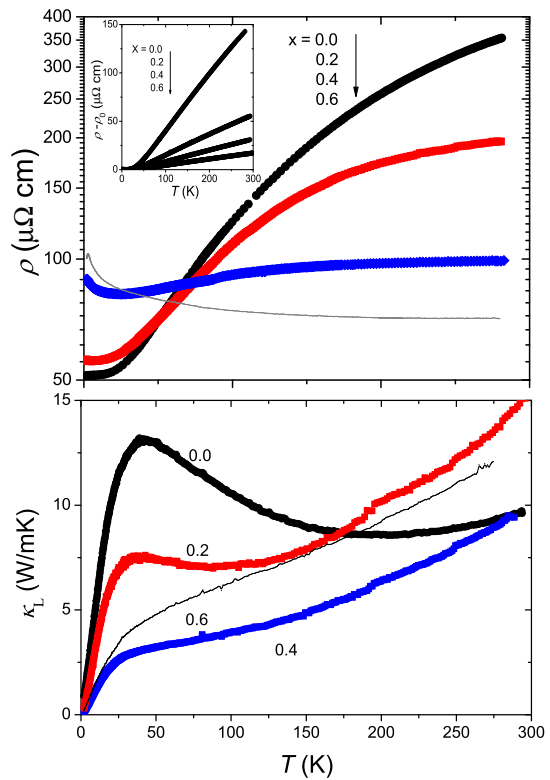}
\caption{The electrical resistivity, $\rho$ (upper panel), and the lattice contribution to thermal conductivity, $\kappa _L$ (lower panel), as a function of temperature for various  
Ce(Ni$_{1-x}$Cu$_x$)$_2$Al$_3$ compounds. For the estimation of $\kappa _{\rm L}$, the Wiedemann-Fran law and the Sommerfeld value of $L$ were employed. Inset shows $\rho$$-$$\rho _0$$(T)$ 
of a reference system, La(Pd$_{1-x}$Ag$_x$)$_2$Al$_3$, which follows the bloch-gr\"uneisen function typical of normal metals with a remarkable reduction of the overall values with increasing 
$x$. The residual resistivity $\rho _0$ of the reference compounds is 3.9, 28.9, 18.8, and 17.2 $\mu\Omega$\,cm for $x$ = 0.0, 0.2, 0.4 and 0.6, respectively. } 
\label{kappa}
\end{figure}

Upon increasing $x$ from 0.0 to 0.6, the electrical resistivity $\rho$ (Fig.~\ref{kappa}, upper panel) evolves from a monotonically increasing to a monotonically decreasing function of 
temperature, again pointing to the decrease of $T_{\rm K}$.  The broad maximum in $\rho(T)$, another estimate of $T_{\rm K}$, shifts from $\sim$200 K for $x$ = 0.2 to  $\sim$100 K for $x$ = 
0.4. Remarkably, $\rho$ above 70 K exhibits a systematic reduction with increasing $x$,  in spite of the increasing residual resistivity $\rho _0$. In combination with the enhancement of 
$S_{\rm max}$, this indicates a simultaneous optimization of $\rho$ and $S_{\rm max}$ that is impossible in classical metals.

The overall resistivity of a non-magnetic, isoelectronic reference system La(Pd$_{1-x}$Ag$_x$)$_2$Al$_3$ \cite{Cudope} exhibits behaviors characteristic of a normal metal 
and decreases steadily with increasing $x$ (Inset of Fig.~\ref{kappa}). For instance, $\rho$ at room temperature changes from 150 $\mu\Omega$\,cm for $x$ = 0.0 to 35 $\mu\Omega$\,cm for $x$ 
= 0.6, reflecting the electronic-structure change in the conduction band. 
The evolution of $\rho$ in the reference system hints that the decrease of $\rho$ with $x$ in Ce(Ni$_{1-x}$Cu$_x$)$_2$Al$_3$ shares the same origin: it is due to a decrease of the usual 
electron-phonon scattering term $\rho _{\rm p}$, assuming that $\rho$ = $\rho _{\rm 0}$+$\rho _{\rm p}$+$\rho _{\rm m}$, where $\rho _{\rm 0}$ is the residual resistivity and $\rho _{\rm m}$ 
the magnetic part occurring only in the Ce compounds.  
On the other hand, $\gamma$ diminishes with $x$ in La(Pd$_{1-x}$Ag$_x$)$_2$Al$_3$ \cite{sun2}, indicative that the full-filled $d$-band of Cu triggers a shift of the Fermi level away from 
the enhanced DOS of $d$ character. Similar effect is confirmed both experimentally and theoretically in Ce(Pd$_{\rm 1-x}$Cu$_{\rm x}$)$_2$Si$_2$ \cite{CPCS}.  
The shift of conduction band from $d$- to $s$- or $p$-like bands can naturally account for the decrease of $\rho$, as well as the decease of $T_{\rm K}$ ($\sim$\,exp($-$1/DOS$(\epsilon _{\rm 
F})J$), as a function of $x$ in Ce(Ni$_{1-x}$Cu$_x$)$_2$Al$_3$. 
$\rho(T)$ for $x$ = 0.4 shows an upturn below 20 K as well as a broad hump at 100 K. This is a sign of the crystal electric field splitting effect on the six-fold-degenerate Ce$^{3+}$ state, 
indicating that not only the ground Kramers doublet, but also one excited multiplet are perhaps involved in the transport properties. The ground-state degeneracy was only recently recognized 
to be important for TE properties in Kondo systems \cite{dege}.

The lattice thermal conductivity $\kappa _{\rm L}$ in CeNi$_2$Al$_3$ (Fig.~\ref{kappa}, lower panel) is characterized by a maximum at $T$ $\approx$ 40 K, a fingerprint of crystalline solid. 
The $\kappa _{\rm L}$ maximum smears out with substituting Cu, leading to a largely reduced $\kappa _{\rm L}$ for $x$ = 0.4 over the whole temperature range.  The reduction of $\kappa _{\rm 
L}$ is attributed to the substitutional chemical disorder that is frequently employed in modern TE development. To better understand the evolution of $\kappa _{\rm L}$ with $x$ in the 
present system, however, it is necessary to consider another phonon-scattering process by, i.e., valence fluctuation, which was recently shown to be an effective way to reduce $\kappa _{\rm 
L}$ in HF or IV systems \cite{ph-vf}. The decrease of Kondo temperature $T_{\rm K}$ with $x$ in the present system evidences the decrease of phonon-scattering rate by valence fluctuation, 
which gives the reason why $\kappa _{\rm L}$ of $x$ = 0.6 is higher than that of $x$ = 0.4.  Furthermore, it should be noticed that the increase of $\kappa _{\rm L}$ with temperature for the 
substituted compounds is not typical of a crystalline solid, and instead is reminiscent of the typical $\kappa _{\rm L}$ observed in amorphous systems.   

\begin{figure}[tp]
\includegraphics[width=0.85\linewidth]{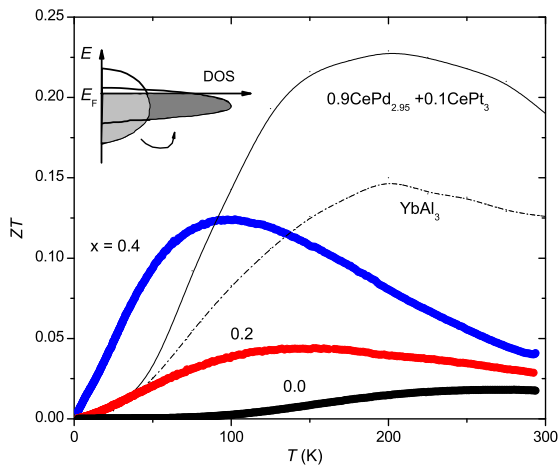}
\caption{The dimensionless thermoelectric figure of merit $zT$ of Ce(Ni$_{1-x}$Cu$_x$)$_2$Al$_3$, in comparison to that of optimized CePd$_3$ and YbAl$_3$ \cite{phyT}. Inset illustrates the 
evolution of the DOS of heavy quasiparticles at near the Fermi level from IV to HF state, which in principle accounts for the enhancement of $S$ in the present Ce(Ni$_{1-x}$Cu$_x$)$_2$Al$_3$ 
system. 
\label{zT}}
\end{figure}

The simultaneous optimization of  $S$, $\rho$, and $\kappa _{\rm L}$ realized in the present substitutional Ce(Ni$_{1-x}$Cu$_x$)$_2$Al$_3$ leads to a systematic, substantial enhancement of 
$zT$ values, as well as a shift of the $zT$ maximum to lower temperature (Fig.~\ref{zT}). In particular at $T$ = 100 K,  $S$ is enhanced by a factor of 4.3 by varying $x$ from 0.0 to 0.4, 
together with a reduction of $\rho$ and $\kappa _{\rm L}$ by a factor of 1.5 and 2.7, respectively (Fig.~\ref{gamma}, lower panel). At this temperature $zT$ for $x$ = 0.4 assumes its peak 
with $zT_{\rm max}$ = 0.125, amounting to half of the maximum $zT$ value observed in optimized CePd$_3$ \cite{phyT}. Much higher $zT$ than that of CePd$_3$ is observed at below 100 K in the 
present system. Comparing the $zT$ values of $x$ = 0.4 to that of $x$ = 0.0 ($zT_{\rm max}$ $<$ 0.02 at 280 K) reveals that the enhancement of $zT$ is considerably significant.  Noticeably, 
the most important among our observations is that the enhancement of $S_{\rm max}$ concomitant to the reduction of $T_{\rm K}$ is not specific to the present system. This is a general 
expectation for HF or IV systems without magnetic ordering \cite{HFTEP},  and has been confirmed in different systems such as Ce(Ni$_x$Pd$_{1-x}$)$_2$Si$_2$ \cite{huo}, though not as 
apparent as in  Ce(Ni$_{1-x}$Cu$_x$)$_2$Al$_3$.

An independent control of $S$, $\rho$ and $\kappa _{\rm L}$ has been theoretically shown to be feasible by confining electrons into a low-dimensional system, and properly fabricated 
thin-film superlattice or nanowire indeed show enhanced thermoelectric properties \cite{lowD}.
By contrast, the simultaneous optimization of the three interdependent thermoelectric parameters in Ce(Ni$_{1-x}$Cu$_x$)$_2$Al$_3$, in particular of $S$ and $\rho$,  is owing to the local 
interaction between the $f$ and conduction bands. The substitution of Cu for Ni in CeNi$_2$Al$_3$ modifies the $d$-like conduction band with enhanced DOS to $s$- or $p$-like bands with 
charge carriers that are  much more mobile. This modification on conduction band leads to not only an increase of electrical conductivity but also an enhancement of the Kondo resonance peak 
in the DOS of the heavy quasiparticles (inset of Fig.~\ref{zT}) due to reduction of $T_{\rm K}$, the latter being crucial for the enhancement of $S$ that measures the energy derivative of 
DOS at the Fermi level. Further enhancement of the thermoelectric figure of merit in Ce(Ni$_{1-x}$Cu$_x$)$_2$Al$_3$ seems achievable through i) substitution with heavier atoms instead of Cu, 
such as Ag, to further reduce $\kappa _{\rm L}$, or ii) approaching the best substitution amount in the vicinity of the critical concentration, $x$ = 0.4. 
Along the same lines, we also stress the significance to study other Kondo compounds which already exhibit moderately enhanced $zT$, for example, CePd$_3$ or systems with large ground-state 
degeneracy. 
It should be reminded that for cryogenic application a comparable efficiency as for high-temperature application is not necessarily crucial; a moderately enhanced $zT$ might make a 
significant impact in this field.  Furthermore, in identification of an appropriate starting Kondo system and the best elements for substitution, theoretical prediction may play an important 
role. For instance, band-structure calculation with respect to a specific Kondo system could predict the suitable elements that might give rise to an ideal TE optimization as observed in the 
present system.

We thank N. Oeschler for helpful discussions.


\begin{thebibliography}{99}
\bibitem{qct} This type of phase transition occurring at zero temperature is usually referred to as quantum critical transition.
\bibitem{Cudope} The corresponding Cu-substituted non-magnetic system LaNi$_2$Al$_3$ was not studied. Instead, replacement of Pd in the iso-structural and iso-electronic system 
LaPd$_2$Al$_3$ by Ag and Cu reveals the similar effects, i.e., reduction of resistivity and $\gamma$ with increasing Cu/Ag concentration.
\end{thebibliography}
\end{document}